\documentclass[runningheads]{llncs}

\usepackage{dialogue}
\usepackage{tabularx}
\usepackage{float}
\usepackage{amsmath}
\usepackage{subcaption} 
\usepackage[T1]{fontenc}
\usepackage{graphicx}
\usepackage{hyperref}

\begin{document}
\title{Optimizing Mastery Learning by Fast-Forwarding Over-Practice Steps}
\titlerunning{Optimizing Mastery Learning by Fast-Forwarding}

\author{Meng Xia\inst{1}\textsuperscript{*}\orcidID{0000-0002-2676-9032} \and
Robin Schmucker\inst{2}\textsuperscript{*}\orcidID{0000-0002-5275-3608} \and
Conrad Borchers\inst{2}\textsuperscript{*}\orcidID{0000-0003-3437-8979} \and
Vincent Aleven\inst{2}\orcidID{0000-0002-1581-6657}}
\authorrunning{M. Xia et al.}

\institute{Texas A\&M University\\
\email{mengxia@tamu.edu}\\
\and
Carnegie Mellon University\\
\email{\{rschmuck,cborcher,aleven\}@cs.cmu.edu}\\
*Equal contribution.}

\maketitle 

\begin{abstract}
Mastery learning improves learning proficiency and efficiency. However, the overpractice of skills--students spending time on skills they have already mastered--remains a fundamental challenge for tutoring systems. Previous research has reduced overpractice through the development of better problem selection algorithms and the authoring of focused practice tasks. However, few efforts have concentrated on reducing overpractice through step-level adaptivity, which can avoid resource-intensive curriculum redesign. We propose and evaluate Fast-Forwarding as a technique that enhances existing problem selection algorithms. Based on simulation studies informed by learner models and problem-solving pathways derived from real student data, Fast-Forwarding can reduce overpractice by up to one-third, as it does not require students to complete problem-solving steps if all remaining pathways are fully mastered. Fast-Forwarding is a flexible method that enhances any problem selection algorithm, though its effectiveness is highest for algorithms that preferentially select difficult problems. Therefore, our findings suggest that while Fast-Forwarding may improve student practice efficiency, the size of its practical impact may also depend on students' ability to stay motivated and engaged at higher levels of difficulty.
\end{abstract}

\begin{figure*}[htpb]
	\centering
 	\includegraphics[width=\textwidth]{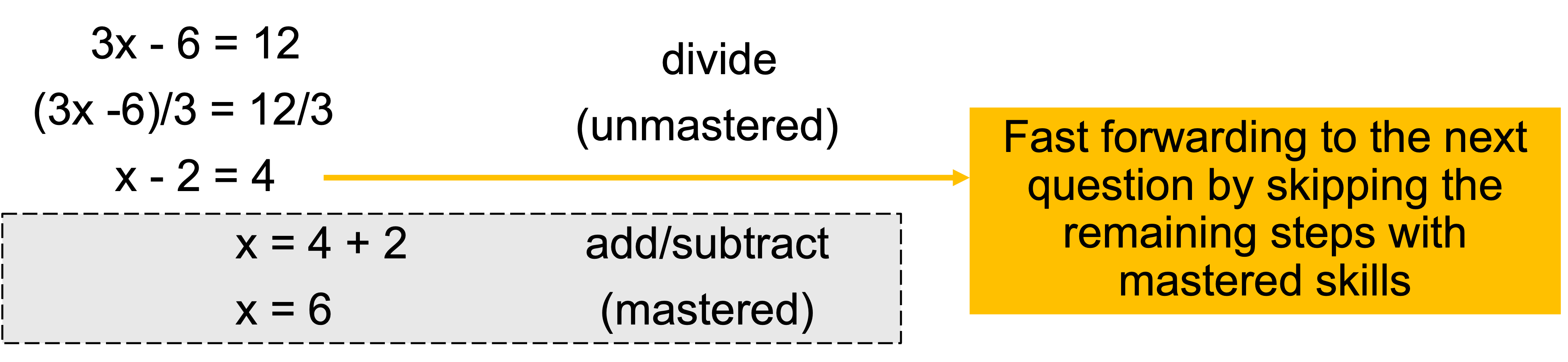}
	\caption{Fast-Forwarding reduces overpractice by avoiding problem-solving paths students have fully mastered. Linear equation-solving steps in gray are known to be mastered by the student and can be fast-forwarded to reduce overpractice.}
	\label{fig:Fast-Forwarding}
\end{figure*}

\section{Introduction}

The field of technology-enhanced learning has successfully optimized problem selection strategies and improved learning efficiency through methods such as Bayesian Knowledge Tracing (BKT) \cite{corbett1994knowledge}. BKT, one the most widely used knowledge tracing algorithms, has proven effective in modeling student knowledge acquisition, enabling tutoring systems to adaptively select practice problems with relevant and adequately difficult skills. In addition, several extensions to BKT have been proposed, including deep learning- (e.g.,~\cite{piech2015deep,zhang20243dg} and regression-based approaches such as AFM, PFA and AugmentedLR (e.g.,~\cite{pavlik2009performance,Schmucker2021:Assessing}).

Adaptive problem selectors are informed by knowledge tracing and grounded in mastery learning \cite{bloom1987response}. Mastery learning assumes that learning and instruction can be divided into atomic skills that students acquire by exercising distinct cognitive operations and receiving feedback from which they learn. This idea aligns with large-scale evidence from several datasets \cite{koedinger2023astonishing} and has improved student performance at scale \cite{ritter2016mastery}. Mastery learning underlies the design of many contemporary tutoring systems and the idea that students require several practice repetitions--learning opportunities--to master each skill. On average, they require seven \cite{koedinger2023astonishing}. However, when tasks or activity sequences are poorly designed, practice activities might lead to \textit{overpractice} and \textit{underpractice} of individual skills.

Overpractice occurs when students practice skills they have already mastered, which is undesirable for learning efficiency if students have not yet mastered other skills \cite{cen2007over}. Overpractice typically occurs when students practice static problem sets where problems include multiple skills, and there is little variability in what skills the problems target. In contrast, underpractice can be due to limited problem pools, meaning students do not master content upon completing a problem set or have insufficient time to practice. While curricular redesign based on learning analytics has been suggested as one method to mitigate over- and underpractice \cite{cen2007over}, avoiding both remains a fundamental challenge in adaptive learning, with learning analytics frameworks on the data-driven redesign of adaptive learning systems considering problem selection a core dimension for design improvement \cite{huang2021general}. In particular, the consequences of reducing overpractice on learning are known to be positive: empirical research has shown that reducing overpractice in CognitiveTutor geometry decreased practice time without impacting learning outcomes \cite{cen2007over}, and such reductions have thus far required a careful and labor-intensive curricular redesign. In the present study, we study whether overpractice can be reduced through more automated means, based on optimization of step-level adaptivity (``inner loop'' \cite{aleven2016instruction}) and having the system take over the mastered steps under some conditions. Such optimization balances the allocation of repetitions, giving students time to master challenging skills.

One reason why overpractice is an open problem is that serving practice problems with a single skill is not a viable solution, as multi-step problems give learners important opportunities to learn, most notably when practicing two skills together is more challenging than practicing them in isolation \cite{heffernan2022developmental}. Hence, it is important to develop novel ways for students to master the skills required to learn while optimizing practice time \textit{and} retaining multi-step problem solving.

An emerging approach to improve adaptive problem selection efficiency is called \textit{focused practice}. This approach involves isolating specific skills that the learner currently lacks. Recent problem selection algorithms consider the overall redundancy in mastery of knowledge and the degree of isolation of unmastered skills in problem selection \cite{huang2018learner}. However, this approach still necessitates the availability of a diverse pool of problems with sufficient breadth and variation in skill density. Additionally, the requirement of designing focused tasks that exercise a limited number of skills necessitates additional content authoring time.

Can a general-purpose, problem pool agnostic, and selector agnostic method for focused practice increase student learning efficiency as measured in overpractice? We propose Fast-Forwarding (Fig. \ref{fig:Fast-Forwarding}), which allows learners to complete multi-step problems up to the point where all remaining problem-solving steps are mastered, thereby contributing only to overpractice. Compared to prior approaches to focused practice \cite{huang2021general,cen2007over}, Fast-Forwarding \textbf{neither requires a specific problem selection mechanism nor depends on curricular redesign} while retaining multi-step problem solving. We quantify the effectiveness of Fast-Forwarding across various contemporary problem selection algorithms in intelligent tutoring systems through simulation studies. We ask: 

\begin{itemize}
    \item[] \textbf{RQ1}: How does learning efficiency as measured in overpractice benefit from Fast-Forwarding in a mastery problem selector given limited practice time?
    \item[] \textbf{RQ2}: How much overpractice does Fast Forwarding reduce until content mastery across different problem selection regimes?
\end{itemize} 

We conducted experiments through simulations based on solution pathways and parameters estimated from open-source data system. We chose equation solving as a domain due to its recursive nature: foundational skills are often repeated at the end of solving for X. However, we note that the \textbf{methods studied here apply to numerous other domains of instruction with step-level problem solving, for which tutoring systems are available} \cite{aleven2009new}. In STEM problem solving, science domain decisions tend to happen first (e.g., selecting physics principles), while later steps involve algebra or arithmetic, which many students may have mastered. In all of these domains, fast-forwarding could remove overpracticed steps by completing mastered steps on behalf of the learner.

\section{Related Work}

\subsection{Learner Models and Problem Selectors} 

Various learner models have been developed to model students' learning and problem-solving behaviors. Bayesian knowledge tracing, arguably the most widely used model in adaptive learning systems today, models learners' latent knowledge of distinct skills as a probability of knowing each~\cite{corbett1994knowledge}. Other dynamic probabilistic models have been proposed to model the dependency of different learning concepts. For example, Learning Factor Analysis~\cite{cen2006learning} models learner knowledge states via logistic regression to deal with learning concepts at different levels, and Performance Factor Analysis~\cite{pavlik2009performance} fits distinct learning rates for correct and incorrect responses. Other models, such as deep knowledge tracing, use neural network architectures to model and predict learner performance~\cite{piech2015deep}. These learner models assess learners' mastery of various skills and how they evolve with provided practice opportunities.

Many problem selection algorithms use estimated content mastery in adaptive learning in diverse ways. Contemporary mastery problem selection algorithms used in Cognitive Tutor typically select the problem with the most unmastered skills (mastery hard)~\cite{aleven2009new,koedinger2010avoiding}. Problem selection algorithms in TutorShop \cite{aleven2025integrated} also offer a similar regime that selects problems with the least, but at least one, mastered skills (mastery easy). However, all these selectors, while generally reducing student practice time \cite{ritter2016mastery}, are limited in that they require a static pool of content and skills. Few notable exceptions exist, where tutoring systems have been designed to adaptively take over specific, commonly overpracticed steps~\cite{ritter1995calculation,koedinger2001cognitive}. Other researchers proposed to supplement problem sets with focused tasks to reduce overpractice, which are problems that target a few skills simultaneously. They particularly designed practice problems that involve only a focused set of skills that students have not mastered and then selected these focused tasks with higher probability~\cite{huang2018learner,huang2021general}. However, designing focused tasks is time-consuming and requires content authoring. Fast-Forwarding, the method proposed in this study, in contrast, can be applied to any existing question pool and dynamically monitors whether certain steps can be skipped based on mastery estimates to mitigate overpractice.

\subsection{Learner Simulation}

Simulating learner data to study learning processes and outcomes is a core method in the field, particularly where obtaining real-world data is challenging, expensive, or impractical \cite{Kaser2024simulated}. These simulations leverage cognitive learning theories to generate performance patterns that can be used to refine instructional methods. For instance, past research has employed simulated learners to identify skill models based on performance patterns, which were subsequently validated on real learner data, leading to improved instruction tailored to these identified skills \cite{li2013general}. Similar refinements have been applied to analyzing error types \cite{weitekamp2020investigating}. Recent research using Generative Adversarial Networks (GANs) has also been applied to enhance the assessment of student knowledge through simulation: Zhang et al. \cite{zhang20243dg} augmented log data from tutoring systems using GANs, which improved the reliability of knowledge estimates.

However, despite these advances in simulations for refining knowledge tracing and instruction, there is a lack of simulated learner experiments that quantify the impact of problem selection algorithms on practice time and efficiency. Previous work has primarily focused on simulating the effects of setting mastery thresholds for error types \cite{fancsali2013optimal} or on performance prediction given different mastery levels \cite{pardos2013towards}. In addition, validating simulated learners on outcomes of interest remains a key research challenge. According to a recent literature review by K\"aser and Alexandron \cite{Kaser2024simulated}, a major issue in previous research using simulated learner data is the lack of systematic studies on the validity of synthetic data. However, there are exceptions, such as the work that used simulated student data in tutoring systems to predict real student performance \cite{matsuda2007predicting}. Similarly, recent studies have increasingly used synthetic data to improve predictive models and predictive validity \cite{zhang20243dg}. Our research simulates the practice efficiency of different problem-selection strategies in tutoring systems. We incorporate empirical estimates of learning rates and solution pathways into our simulations to improve validity.

\section{Methods}

\subsection{Study Context and Dataset}
\label{sec:method:data}

We utilize an open-source dataset from APTA, a collaborative tutoring system for middle school equation-solving, available on PSLC DataShop (datasets \#5549 and \#5604) \cite{koedinger2015data}. The study sample consists of data from IRB-approved classroom studies conducted in math classes at two public suburban middle schools in a mid-sized city in the eastern U.S. Overall, 164 6th- to 8th-grade students engaged with a series of multi-step linear equation-solving problems \cite{borchers2024combining}.

We analyzed the data comprising 10,937 entries (N = 10,937) of completed and timestamped problem-solving steps using the DataShop student step rollup tool \cite{koedinger2015data}. The log data provide a detailed description of the student solution paths, meaning the order of operations used to solve individual problems. This allows for a fine-grained analysis of skills (e.g., add constant) students employ to derive solutions, offering deeper insights into their step-level problem-solving processes and serving as reference pathways for our simulation studies.

We only considered students' first attempts, as they reflect the students' initial response to solving linear equation steps without tutor assistance. Steps exhibiting hint usage were classified as incorrect attempts~\cite{koedinger2015data}.

\subsection{Computation of Overpractice}

We define overpractice on a skill as the number of practice opportunities (e.g., problem-solving steps) on a skill given a predicted error rate on that skill below 0.05, in line with the mastery criteria in the Lynnette tutor \cite{long2018exactly}. Table \ref{tab:empirical_analysis} reveals several key findings regarding student performance and practice behavior across different skills. \textbf{This analysis served as a baseline for how much overpractice was present in our dataset from real students.} Simulations, as described below, were then used to determine how much students could potentially save if they had unlimited practice time and under different, alternative problem-solving regimes.

\begin{table*}[htpb] %
    \centering
    \setlength{\tabcolsep}{7pt}
    \caption{Student mastery rates, practice and overpractice opportunities for each skill based on student data. The first row shows the \% of students who mastered each skill, the second row the average number of problem-solving steps per skill, and the third row the number of steps taken after mastery (overpractice).}
    \resizebox{1.00\textwidth}{!}{\begin{tabular}{cccccccccc}
    \hline
    &
    \parbox[t]{1cm}{cancel-const} & 
\parbox[t]{1cm}{cancel-var} & 
\parbox[t]{1cm}{comb-const} & 
\parbox[t]{1cm}{comb-var} & 
\parbox[t]{1cm}{division-complex} & 
\parbox[t]{1cm}{division-simple} & 
\parbox[t]{1cm}{simplify-division} & 
\parbox[t]{1.3cm}{add/subtr-const} & 
\parbox[t]{1.3cm}{add/subtr-var} \\  
    \hline
    Mastered (\%)      & 74.4 & 6.7 & 73.8 & 6.7 & 26.8 & 79.3 & 79.3 & 68.9 & 6.7 \\
    Avg Opps.          & 15.9 & 1.1 & 16.2 & 0.9 & 4.1  & 15.4 & 18.3 & 16.3 & 1.2 \\
    Avg Overpr.        & 6.4  & 0.5 & 6.7  & 0.5 & 1.5  & 7.8  & 9.8  & 6.2  & 0.5 \\
    \hline
    \end{tabular}}
    \label{tab:empirical_analysis}
\end{table*}

\textbf{Varied Skill Mastery and Practice Engagement}
The ``Mastered'' column indicates the proportion of students who achieved mastery for each skill, while the ``Avg Opps'' column denotes the average number of completed practice opportunities per skill.
Average practice engagement varies significantly, with some skills having a higher average number of completed practice opportunities than others, such as \textit{division-simple}.

\textbf{Severe Overpractice}
The column ``Avg Overpr.'' delineates the average number of practice opportunities associated with overpractice for each skill. A considerable number of students experience overpractice across all skills.
While more students overpractice the skills, some students achieve mastery of them.

\subsection{Fast-Forwarding}

\textbf{Fast-Forwarding} is a novel technique introduced in Fig. \ref{fig:Fast-Forwarding} that terminates practice problems early and forwards students to the next problem during the active problem-solving process (``inner loop'' \cite{Vanlehn2006:Behavior}). Fast-Forwarding is a flexible technique that can augment existing problem selection algorithms. Its only requirement is access to a pool of multi-step practice problems--a precondition for any step-level adaptation. 

Fast-Forwarding optimizes practice efficiency by focusing on problem-solving steps that are unmastered, allowing students to achieve proficiency more quickly without redundant practice of already mastered skills (i.e., overpractice). However, a key consideration was to exclude skills from the solution pathway \textbf{if and only if all remaining skills had already been mastered}. This decision is informed by the literature on the composition effect \cite{heffernan2022developmental}, which is explained below.

Specifically, we did not adopt a model that would fast-forward students through problem-solving steps for \textit{any} skill they had already mastered in the middle of the problem. Advancing students this way might limit the benefits of practicing combinations of skills in multi-step problem solving; in that case, unmastered in combination with mastered skills. This effect, known as the composition effect, suggests that a skill can be more challenging when practiced alongside a second skill and may require integrated practice \cite{heffernan2022developmental}. Thus, Fast-Forwarding ensures students continue to practice skill combinations until all involved skills are mastered, preventing premature advancement that could leave important compositions unpracticed or systematically underpracticed.

\subsection{Simulation Methodology}
\label{sec:methods:sim}

We simulate student performance (AFM; \cite{cen2006learning}) and trace student knowledge (BKT; \cite{corbett1994knowledge}) through two distinct models. AFM and BKT work together to simulate student responses and different problem-selection regimes. Our simulations focus on problem-solving processes at the step-level, in line with AFM \cite{cen2006learning} and prior analyses of our study data \cite{borchers2024combining}. Besides the present description, we promote method adoption and reproducibility by open-sourcing our code on GitHub\footnote{\href{https://github.com/rschmucker/fast-forwarding}{https://github.com/rschmucker/fast-forwarding}}.

\textbf{Knowledge Tracing:} Mimicking the knowledge tracing within the Lynnette tutoring system, we employ BKT~\cite{corbett1994knowledge} to analyze student responses and consider mastery to be achieved at a threshold of 95\%, which is the threshold in our study data's Lynnette tutoring system \cite{long2018exactly}. BKT models student learning as a hidden Markov process where each skill is represented by a binary latent state (learned or unlearned). The model updates the probability of mastery based on observed correct and incorrect responses, using four key parameters: prior knowledge ($p_{init}$), the initial probability that a student knows the skill before any practice, learning rate ($p_{learn}$), the probability of transitioning from the unlearned to the learned state on each opportunity, guess rate ($p_{guess}$), the probability of answering correctly by guessing when the skill is unlearned, and slip rate ($p_{slip}$), the probability of answering incorrectly due to a mistake when the skill is learned. Given an observed response, BKT applies Bayes' rule to update the posterior probability of mastery. If a student answers correctly, the model increases the mastery probability, accounting for possible guessing. If they answer incorrectly, it decreases the probability, adjusting for slips. Mastery is typically declared when the posterior probability of a student being in the learned state exceeds a predefined threshold (here, 95\%). For our simulations, we used standard TutorShop BKT parameters for all skills ($p_{init}$ = 0.25, $p_{learn}$ = 0.2, $p_{guess}$ = 0.2, $p_{slip}$ = 0.1), ensuring consistency with typical cognitive tutor implementations \cite{aleven2025integrated}. For this study, we simulate $10{,}000$ simulated students that exhibit a normal distribution in initial AFM proficiency levels. This sample size is expected to yield sufficiently reliable results. We note that statistical significance testing our context is not meaningful, as more simulation samples will always tend to produce significant differences, no matter the minuteness of effect sizes.

\textbf{Student Simulation:} We simulate student learning data using AFM~\cite{cen2006learning}, a logistic regression model of knowledge acquisition (represented as the probability of getting a specific problem-solving step requiring specific skills correct) as a function of learners' initial proficiency levels, skill-specific difficulty, and skill-specific learning rates. To promote the validity of our methodology, we fit the standard AFM model \cite{cen2006learning} to our open source dataset (see Section \ref{sec:method:data}) to obtain estimates of student acquisition rates across different skills, which then serve the simulation of performance in our experiments. Skill-level learning rates used for simulation are encoded in interactions between the learning opportunity count and the skill exercised at a given problem-solving step. This AFM learning rate should not be confused with the BKT parameters for knowledge tracing serving problem selection, where we use standard default settings from Lynnette, which assume constant skill learning rates across all skills (see above) to mimic student experiences in the original study. 

We define prototypical solution paths for each practice problem based on the most frequent skill-application path for each problem in our data. At each step in the simulation, we sample a student response based on the correctness probability estimated by the AFM model. We update the AFM simulation and BKT models before moving to the next step or practice problem. In line with Lynnette \cite{long2018exactly}, BKT mastery estimates provide a basis for all problem-selection algorithms--including the proposed Fast-Forwarding method. To address RQ2, we implement a mechanism that replenishes the problem pool after it has been exhausted, meaning that the problem pool is reset to its initial state. This mechanism ensures that the simulation only terminates once all skills are mastered.

\textbf{Problem-selection Algorithms} Fast-Forwarding offers a flexible method for enhancing the efficiency of existing problem-selection algorithms. To verify its capabilities, this study integrates Fast-Forwarding into diverse problem selectors: 

\begin{itemize}
    \item[] \textbf{Random}: Selects problems from the pool at random without replacement.
    \item[] \textbf{Deterministic}: Selects problems following a domain expert-defined order taken directly from the standard Lynnette set suite \cite{long2018exactly} designed to surface increasingly difficult and complex problems over time. 
    \item[] \textbf{Mastery Easy}~\cite{aleven2009new}: Employs student mastery estimates to select problems by averaging the difficulty of all available opportunities and serving the problem with the \textit{lowest} difficulty based on mastery estimates while ensuring that the problem has at least one unmastered skill.
    \item[] \textbf{Mastery Hard}~\cite{aleven2009new}: Employs student mastery estimates to select problems by averaging the difficulty of all available opportunities and serving the problem with the greatest difficulty, usually featuring the most unmastered skills. This selector is the standard problem selector in Lynnette \cite{long2018exactly} and is evaluated in RQ1.
    \item[] \textbf{Focused Practice}~\cite{huang2018learner}: More recent selection algorithm that considers problem difficulty while preferentially sampling problems with few skills via multinomial sampling.
\end{itemize}

\section{Results}

\subsection{Learning Efficiency Benefits (RQ1)}

For the legacy mastery-based problem selector, Fast-Forwarding reduced overpractice by 35.7\% across the simulated students, as measured in learning opportunities to apply mastered skills. Assuming a median number of steps, underpractice was not changed by Fast-Forwarding (M = 0.44 for Fast-Forwarding vs. M = 0.43 without). Hence, Fast-Forwarding mitigates overpractice while not introducing more underpractice, meaning that overall mastery levels do not suffer.
\begin{figure}[ht]
    \centering
    \includegraphics[width=\linewidth]{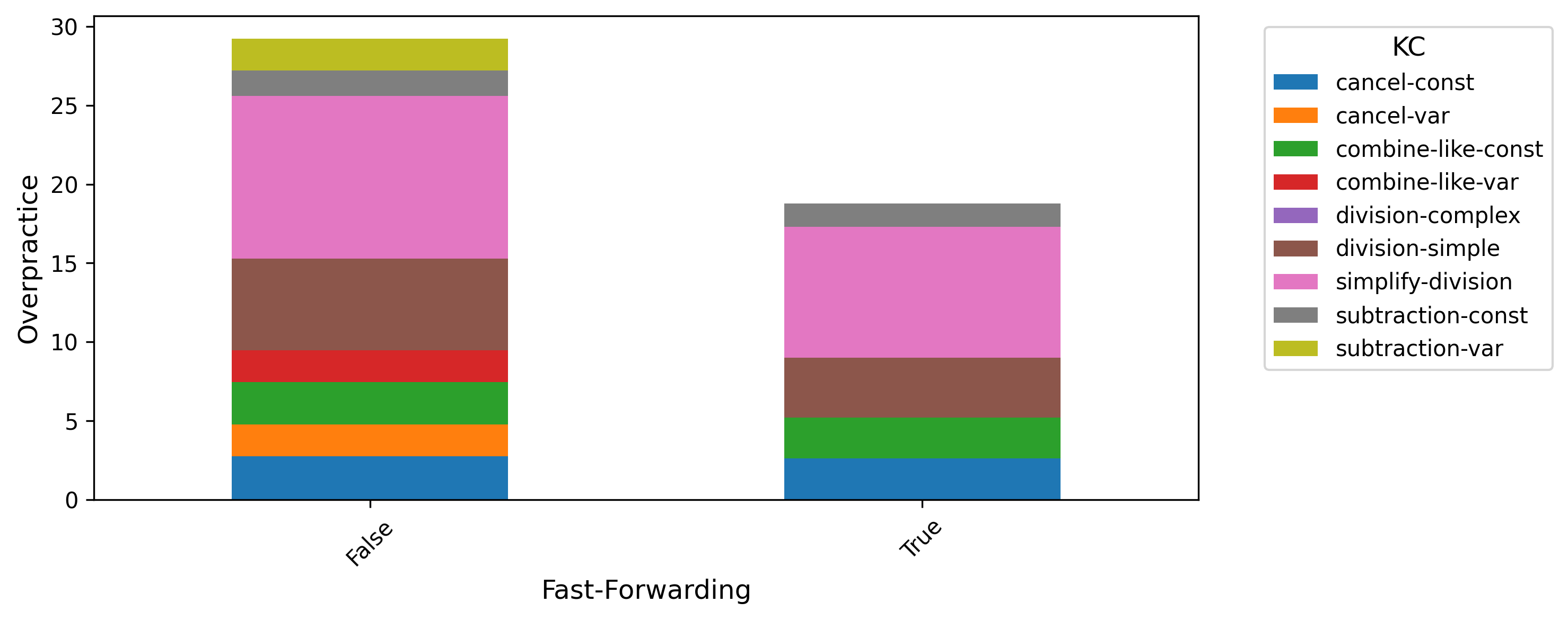}
    \caption{Stacked bar plot showing the reduction in overpractice across individual skills when Fast-Forwarding is applied.}
    \label{fig:stacked_overpractice_kc}
\end{figure}

Fig. \ref{fig:stacked_overpractice_kc} shows the sum of overpractice across various skills averaged across students. Overall, without Fast-Forwarding, overpractice is substantially higher, which can be attributed to continuing practice without early termination of fully mastered pathways. Notably, Fast-Forwarding removes overpractice entirely from some skills. For example, in our sample, Fast-Forwarding removed overpractice from the skills ``subtract variable,'', ``combining like variables,'' and ``cancel variable.'' These are skills that typically appear at the end of practice problems. The same is true for skills involving handling constants. However, these skills may also be required to set up more advanced skills through transformation, such as division. These ``setup'' skills may not always be eliminated by Fast-Forwarding. For example, 3x - 2 = 4 requires first subtracting as division by 3 upfront is not permitted in Lynnette, which only accepts whole numbers \cite{long2018exactly}).

\subsection{Effects Across Problem Selectors (RQ2)}

\begin{figure}[ht]
    \centering
    \includegraphics[width=.9\linewidth]{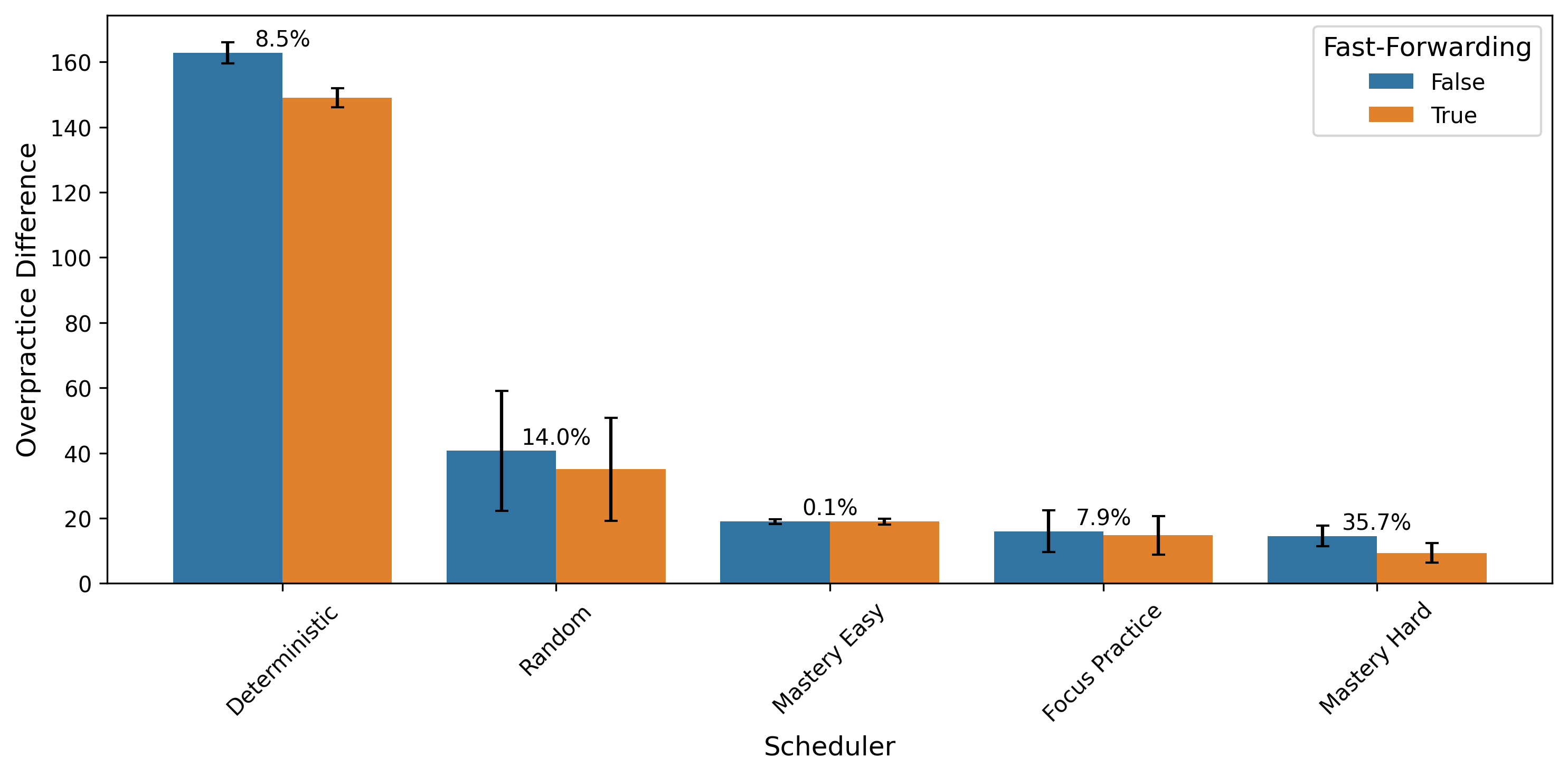}
    \caption{Overpractice difference across problem selectors for Fast-Forwarding, including \% decreases when Fast-Forwarding is applied. Error bars are 2 $SD$ across students.}
    \label{fig:scheduler_overpractice_sorted_high_res}
\end{figure}

One strength of Fast-Forwarding is that it does not depend on the specific problem-selection algorithm used. Fig. \ref{fig:scheduler_overpractice_sorted_high_res} shows the impact of applying Fast-Forwarding on overpractice across various problem selection regimes. This simulation aimed to determine how much overpractice students would encounter to reach mastery, assuming unlimited time. Notably, Fast-Forwarding consistently reduced overpractice across all problem selectors, ranging from 0.1\% to 35.7\%. The largest reduction for the legacy mastery hard problem selector aligns with RQ1 findings, but with the difference that no learning opportunity limit was imposed here, reducing overpractice somewhat more. 

As error bars in Fig. \ref{fig:scheduler_overpractice_sorted_high_res} indicate, between-student variation in overpractice differed across problem selectors. As expected, random scheduling led to the largest variation, and deterministic scheduling led to the second-lowest variation. Despite this variation, Fast-Forwarding still led to a notable reduction in overpractice of 0.67 $SD$ for this problem selector. Surprisingly, mastery easy led to the lowest variation in overpractice and only marginal overpractice reductions from Fast-Forwarding. Notably, the modest reduction in the focused practice problem selector was concurrent with the second-largest variation in overpractice across students (7.9\%, corresponding to a reduction of 0.41 $SD$). We suspect this variation stems from increased randomness in the problem selector, both choosing skills \textit{and} problems randomly. Accordingly, the mastery hard problem selector was more desirable based on the reported largest reduction in overpractice (35.7\%; RQ1) \textit{and} lower between-student variation, corresponding to a substantial effect size of 3.4 $SD$.

Finally, given the content unit being designed to include sufficient problems to allow all students to master content, the deterministic problem selector, requiring all problems, was associated with substantial overpractice, over ten times larger than mastery-based problem selectors. Overall, assuming students exert continued practice effort, mastery is obtained with the least overpractice for problem selectors with Fast-Forwarding, especially those who preferentially serve difficult problems.

\section{Discussion}

Mastery learning makes learners improve efficiently \cite{ritter2016mastery}. Problem selection is key in improving learning analytics applications that support mastery learning, most notably tutoring systems \cite{huang2021general}. The objective of the present study was to evaluate a novel method that promises to improve the efficiency of mastery learning as measured in overpractice \cite{cen2006learning} while preserving multi-step problem solving and its benefits to learning \cite{heffernan2022developmental}. Through simulation based on parameters informed by real student data, we applied our method, Fast-Forwarding, to mitigate overpractice across contemporary and legacy methods for problem selection in a tutoring system.

\subsection{Fast-Forwarding Under Legacy Practice Conditions (RQ1)}

RQ1 focused on learning efficiency gains in a legacy mastery problem selector, prioritizing hard problems given limited practice time. Informing practice time by average step completions of students in an open-source dataset and using learning parameters from the same data, our analysis aimed at closely estimating how much practice students could save by our Fast-Forwarding method. Under limited practice time, our findings indicate that students could save up to about a third of overpractice in a linear equation-solving tutor \cite{long2018exactly} by Fast-Forwarding. In particular, this overpractice benefit was not accompanied by an underpractice deficit, indicating that the overall efficiency of practice improved.

Our findings pose two central questions for future research. First, while Fast-Forwarding is generally applicable to any mastery-based problem selector and content unit, equation solving is a task domain that naturally affords a high level of overpractice: More complex problems require more steps than simple problems, and intermediate problem states are similar to simpler problems, whose problem steps are repeated. Some of these ``foundational steps'' at the end of equations were eliminated from overpractice through Fast-Forwarding, for example, combining variables. However, other task domains might not have such hierarchical problem structures, such that Fast-Forwarding could potentially be less effective for such domains, which merits future research. Specifically, our method contributes a novel method of overpractice reduction in addition to curricular redesign \cite{cen2007over} and problem selection optimization \cite{huang2021general}. It is possible that each of these methods may be most effective across different domains. However, we argue that our method can be most readily applied to any multi-step problem-solving domain with knowledge tracing \cite{aleven2025integrated}, including any problem selector.

Notably, Fast-Forwarding could not eliminate \textit{all} underpractice in equation solving, potentially because some steps are required to set up more complex, unmastered operations, such as division, where students first need to subtract constants (a more foundational skill) to divide adequately (e.g., in 3x - 2 = 4, where division by 3 upfront, leading to non-integers, is not permitted in Lynnette \cite{long2018exactly}). While Lynnette could fast-forward students to the state 3x = 6 to just practice division, dealing with constants, in this case, may pose an important learning opportunity, assuming a composition effect whereby the same skill is more or less difficult depending on co-occurring skills \cite{heffernan2022developmental}. Although Fast-Forwarding currently preserves such learning opportunities, future research may investigate to what extent learners could benefit from intermediate forwarding.

\subsection{Fast-Forwarding is Effective for High-Difficulty Selectors (RQ2)}

Assuming unlimited practice time, RQ2 focused on the effectiveness of Fast-Forwarding flexibly applied to different contemporary and legacy problem selection regimes. We find that, in a content pool with enough problems to allow all students to master all skills, mastery learning as such already substantially reduces overpractice compared to deterministic problem selectors that require learners to fully complete a unit before moving on. However, for all mastery-based algorithms, including recent ones that try to create more opportunities for focused practice and less overpractice \cite{huang2018learner}, Fast-Forwarding further reduced overpractice. However, it was especially effective for selectors that prioritized difficult problems while minimizing overpractice differences between students, which may be desirable for creating practice regimes that are equally effective for all students \cite{koedinger2023astonishing}. One interpretation of this finding is that mastery algorithms that prefer harder problems tend to serve problems with many skills (i.e., the mastery hard algorithm used in this study tends to maximize the number of unmastered skills in a problem). In contrast, the mastery easy algorithm tends to serve problems with fewer skills (with at least one unmastered skill) \cite{koedinger2010avoiding}. It is hence expected that longer problems allow for more opportunities for the Fast-Forwarding mechanism to fast-forward the student, reducing overpractice to a larger extent. Second, selectors that preferred harder problems in our selection also led to the least problem selection overall. This is perhaps because, unlike Mastery Easy, such selectors more heavily punish problems that include mastered knowledge, as Mastery Easy only requires at least one skill in a problem to be unmastered.

\subsection{Considerations Towards Applying Fast-Forwarding in Practice}

While our findings suggest that Fast-Forwarding alongside selecting hard problems reduces overpractice most (based on simulation assumptions about learning based on real student data; a key issue identified in past research \cite{Kaser2024simulated}), it is not self-evident that such problem selection regimes will be most effective for learning in practice. Although mastery learning is effective \cite{ritter2016mastery}, it can be difficult for students to maintain the effort required to engage in practice. Studies published in the 1970s documented that student participation in computer-assisted instruction decreased with difficulties below 80\% expected accuracy and was high when problems were easy \cite{miller_hess_1972}. While motivational theory suggests that a moderate level of difficulty is the most desirable level for motivation, more recent research also suggests that novelty and variation in learning tasks can compensate for some of the adverse effects of high difficulty on motivation \cite{Lomas2014}. Engagement in practice is crucial for the learning benefits of tutoring systems \cite{koedinger2023astonishing}. Therefore, potentially more important than practice efficiency (i.e., much practice with a little overpractice will usually result in more learning than little or no practice with no overpractice), future research is required to determine to what extent Fast-Forwarding can benefit students and their learning most. Although our findings suggest that Fast-Forwarding is especially helpful in reducing overpractice in problem selectors prioritizing hard problems, increasing difficulty (i.e., removing mastered skills) even further through Fast-Forwarding may not always be desirable for student motivation and engagement. Increasing the novelty of tasks to potentially and partially offset this effect \cite{Lomas2014} could be realized through authoring additional focused tasks \cite{koedinger2016closing} or potentially only executing Fast-Forwarding sometimes and at random, which would also lower overall practice difficulty by occasionally including mastered skills. Independently, current mastery thresholds (i.e., 95\%) might be better if they were higher to account for forgetting, an interesting question outside the scope of our paper with some nascent evidence \cite{zhang2025how}. These design considerations, including how to best introduce and realize Fast-Forwarding in the tutoring system interface, are beyond the scope of the present study and subject to future research.

\subsection{Limitations and Future work}

First, our findings are limited to a single domain (i.e., equation solving). Given the portability of Fast-Forwarding to any problem selector and problem domain with empirically derived student solution pathways, our findings merit future research into more domains of tutoring, which our open-source code enables.

Second, our Fast-Forwarding implementation was limited because Bayesian Knowledge Tracing (BKT) parameters were not fitted for individual students, which may have impacted the precision of skill mastery predictions. Although our BKT values used aligned with actual Lynnette settings \cite{long2018exactly}, future research should consider individual fitting BKT parameters to learner data for more precise problem selection and simulation.

Third, our findings are limited to a single skill model, while potential difficulty factors in complex skill models may better describe the learner data \cite{long2018exactly}. A risk of using Fast-Forwarding when applied to coarse-grained skill models is taking away practice from difficult, hidden skills. Although skill model refinement is a key challenge in data-driven tutoring system improvement \cite{huang2018learner}, future research may study what effects Fast-Forwarding has on more coarse-grained skill models.

\section{Conclusion}

This study introduced the generalizable Fast-Forwarding method to enhance mastery learning efficiency by reducing overpractice while preserving the benefits of multi-step problem solving. Our simulation studies showed that Fast-Forwarding reduces overpractice by up to one-third, particularly in selectors prioritizing harder problems. This improvement makes it a promising tool for optimizing tutoring systems, a key application area of technology-enhanced learning. These findings suggest that balancing efficiency with task difficulty, which may reduce student engagement if too high, should be a key consideration in future research to improve learning in practice.

\section*{Acknowledgments}

This research was funded by the Institute of Education Sciences (IES) of the U.S. Department of Education (Award \#R305A220386).

\bibliographystyle{splncs04}
\bibliography{main}

\end{document}